\documentclass[a4paper,12pt]{article}

\usepackage{pstricks}
\usepackage{color}
\usepackage{amssymb,amsmath,bm,bbm}
\usepackage{epsf,epsfig}
\usepackage{afterpage}
\usepackage{longtable}
\usepackage{cite}
\usepackage{latexsym,mathrsfs,dsfont}
\usepackage{graphicx}
\usepackage{url}
\usepackage{paralist}
\usepackage{bbold}
\usepackage{appendix}
\usepackage{hyperref}
\usepackage{marginnote}

\usepackage{soul}

\definecolor{light-gray}{gray}{0.90}
\definecolor{dgreen}{rgb}{0.0, 0.5, 0.0}

\begin{document}

\hfill {\tt CERN-TH-2025-254}

\def\thefootnote{\fnsymbol{footnote}}

\begin{center}
\Large\bf\boldmath
\vspace*{1.cm} 
Probing scalar and pseudoscalar new physics using rare kaon decays
\end{center}
\vspace{0.6cm}

\begin{center}
G.~D'Ambrosio$^{1}$\footnote{Electronic address: gdambros@na.infn.it}, 
A.M. ~Iyer$^{2}$\footnote{Electronic address: iyerabhishek@physics.iitd.ac.in}, F.~Mahmoudi$^{3,4,5}$\footnote{Electronic address: nazila@cern.ch}, 
S. Neshatpour$^{3}$\footnote{Electronic address: s.neshatpour@ip2i.in2p3.fr}\\
\vspace{0.6cm}
{\sl $^1$INFN-Sezione di Napoli, Complesso Universitario di Monte S. Angelo,\\ Via Cintia Edificio 6, 80126 Napoli, Italy}\\[0.4cm]
{\sl $^2$Department of Physics, Indian Institute of Technology Delhi,\\ Hauz Khas, New Delhi-110016, India}\\[0.4cm]
{\sl $^3$Universit\'e Claude Bernard Lyon 1, CNRS/IN2P3, \\
Institut de Physique des 2 Infinis de Lyon, UMR 5822, F-69622, Villeurbanne, France}\\[0.4cm]
{\sl $^4$Theoretical Physics Department, CERN, CH-1211 Geneva 23, Switzerland}\\[0.4cm]
{\sl $^5$Institut Universitaire de France (IUF), 75005 Paris, France }\\[0.4cm]
\end{center}
\renewcommand{\thefootnote}{\arabic{footnote}}
\setcounter{footnote}{0}

\vspace{1.cm}
\begin{abstract}
\noindent

Rare kaon decays provide sensitive tests of new physics. In this work, we focus on scalar and pseudoscalar operators, analysing the $K\to \pi \ell^+\ell^-$ and $K\to \ell^+\ell^-$ decays. We highlight the complementary role of different modes: $K^+\to\pi^+\ell^+\ell^-$, in particular the forward--backward asymmetry in the muon channel as a clean probe of scalar effects, the stringent constraints from $K_L\to \mu^+\mu^-$, and the discovery potential of future measurements of $K_S\to \mu^+\mu^-$ and $K_L\to \pi^0 \ell^+\ell^-$. The interplay between charged and neutral modes underscores the complementarity of NA62, the LHCb upgrade, and KOTO-II. 

\end{abstract}

\thispagestyle{empty}

\clearpage

\section{Introduction}
Rare kaon decays provide a uniquely sensitive area for probing flavour dynamics beyond the Standard Model (SM)\cite{Cirigliano:2011ny,AlvesJunior:2018ldo,Isidori:2010kg,Neshatpour:2024lij}. The extreme suppression of these processes in the SM, together with the availability of precise theoretical control for several key channels, makes them complementary to $B$-physics observables in constraining new physics (NP). 
Most recent progress in recent years has focused on modes dominated by vector and axial-vector operators, most notably $K\to \pi \nu \bar\nu$. By contrast, scalar and pseudoscalar interactions, despite their generic appearance in a wide range of extensions of the SM such as scenarios with extended Higgs sectors, leptoquarks, or light new degrees of freedom, have received comparatively less systematic attention. Their effects can nevertheless be sizeable in several kaon channels, and in some cases provide qualitatively new experimental handles absent in the SM. This motivates a comprehensive reassessment of scalar and pseudoscalar probes in light of the upcoming experimental programme.

Among the most sensitive observables are the dileptonic decays $K\to \pi \ell^+\ell^-$ and $K\to \ell^+\ell^-$, which constitute the most direct probes of such couplings. Their phenomenology is shaped by the interplay of short-distance dynamics and long-distance contributions.
The forward–backward asymmetry, in the muon mode, essentially vanishing in the SM, provides a particularly clean null test of scalar interactions. The purely leptonic decays $K_L\to \mu^+\mu^-$ and $K_S\to \mu^+\mu^-$ offer complementary sensitivity: the former is limited by long-distance uncertainties but already yields tight constraints on pseudoscalar couplings, whereas the latter remains an exceptionally clean probe of new scalar dynamics due to the absence of sizeable interference with the SM. Finally, the mode $K_L\to \pi^0 \ell^+\ell^-$ can provide complementary probes of scalar and pseudoscalar Wilson coefficients.

With the next generation of kaon experiments, namely NA62 extended dataset, LHCb gaining access to large samples of charged and neutral kaon decays, and KOTO-II aiming for unprecedented sensitivity, the exploration of these observables is entering a crucial stage~ \cite{deBlas:2025gyz,DAmbrosio:2023irq}. In particular, the ability of these experiments to measure observables sensitive to different combinations of scalar and pseudoscalar operators offers a powerful strategy to identify or exclude correlated patterns of new physics.

In this work, we present a systematic study of scalar and pseudoscalar contributions to these rare kaon decays, highlighting both the present constraints and the future prospects, and emphasising the complementarity of charged and neutral modes across different experimental frontiers. All the results presented in this work are produced using the \texttt{SuperIso v5.0} program~\cite{Mahmoudi:2007vz,Mahmoudi:2008tp,Mahmoudi:2009zz,Neshatpour:2021nbn}.

This paper is organised as follows. In section 2 we review the current status of scalar operator constraints from the charged mode $K^+\to \pi^+ \ell^+\ell^-$, with particular emphasis on the differential decay rate and the forward-backward asymmetry, following the work done in Ref.~\cite{DAmbrosio:2024rxv}. Section 3 is devoted to the impact of scalar and pseudoscalar interactions on the dileptonic decays $K_{L,S}\to \mu^+\mu^-$, where we analyse two representative new physics scenarios and discuss the resulting correlations between charged and neutral modes. In section 4 we investigate the sensitivity of $K_L \to \pi^0\ell^+\ell^-$ to scalar and pseudoscalar operators and assess the prospects for future measurements at KOTO-II. We summarise our findings and conclude in section 5.

\section{Scalar operators with $K^+\to \pi^+ \ell^+\ell^-$}

We consider simplified extensions of the Standard Model that includes only scalar~($S$) and pseudoscalar~($P$) type operators. Ignoring BSM operators of SM-like structure in favour of $S$ and $P$ operators is well motivated from a model building perspective, where the additional degrees of freedom (spin-1 or spin-1/2) are significantly heavier than only the spin-0 bosons. Prominent examples of such scenarios arise in supersymmetric models with light scalars~\cite{Martin:1997ns,Ellwanger:2009dp,Branco:2011iw} and in composite dynamics with pseudo-Nambu-Goldstone bosons (PNGB's)~\cite{Cacciapaglia:2020kgq}. As a result, additional contributions to the different observables in the kaon sector would be dominated by the scalar and pseudoscalar operators. 
The simplest parametrisation of the effective Hamiltonian with these operators can be written as (see~e.g.~\cite{Buchalla:1995vs,Bobeth:2017ecx}):
\begin{equation}
\mathcal{H}_{\rm eff}=-\frac{4G_F}{\sqrt{2}}\lambda_t\frac{\alpha_e}{4\pi}\sum_k \left( C_k^{\ell}O_k^{\ell} + C_{Q}^{\prime\,\ell}Q_k^{\prime\,\ell}\right)\,+ \text{h.c.}\,,
\label{eq:hamiltonian}
\end{equation}
with $\lambda_t\equiv V^*_{ts}V_{td}$. 
The dominant SM operators contributing to the decays considered in this paper are
\begin{align}\nonumber
&{O}_9^{\ell} = (\bar{s} \gamma_\mu P_L d)\,(\bar{\ell}\gamma^\mu \ell)\,,
&&{O}_{10}^{\ell} = (\bar{s} \gamma_\mu P_L d)\,(\bar{\ell}\gamma^\mu\gamma_5 \ell)\,,
\end{align}
and for scenarios beyond the SM, we have:
\begin{align}
&O_{S}^{\ell} = (\bar{s} P_R d)\,(\bar{\ell} \ell)\,,
&&O_{P}^{\ell} = (\bar{s} P_R d)\,(\bar{\ell} \gamma_5 \ell)\,,
\end{align}
where $\ell$ corresponds to the lepton flavour.
We choose to adopt a minimal flavour violating-like scheme to characterise these additional contributions; translations to other parametrisations can be trivially obtained by appropriate rescaling of the Wilson coefficients. Possible new physics effects are conveniently parametrised as shifts in the Wilson coefficients with respect to their SM values, $C_i = C_i^{\rm SM} + \delta C_i$. For the scalar and pseudoscalar Wilson coefficients, the corresponding SM contributions are negligible. Furthermore, in this work we do not take into account the chirality-flipped primed operators.

\subsection{Bound on the scalar Wilson coefficient}
In light of the ongoing measurements at NA62, this mode holds the potential to offer the main insight into the scalar operators from kaon physics. In the Standard Model, its decay width is characterised by a dominant long-distance (LD) vector contribution. In the context of Eq.~(\ref{eq:hamiltonian}), the scalar operator can also contribute directly, as well as by means of interference with the vector contribution.
In terms of the vector and scalar form factors, $f_V$ and $f_S$, respectively, the amplitude for this process can be written as~(see e.g. \cite{E865:1999ker,Chen:2003nz,Gao:2003wy,DAmbrosio:2024rxv}): 
\begin{align}
{\cal M} = \frac{\alpha G_F}{4\pi} f_V(z)\, P^\mu \bar{\ell}\gamma_\mu \ell + G_F M_K f_S \, \bar{\ell} \ell \,,
\end{align}
where the dilepton invariant mass is given by $q^2 = z M_K^2$.
From this, the differential decay spectrum ($d\Gamma/dz$) reads as:
\begin{align}\label{eq:dGdz}
\frac{d\Gamma}{dz} &=  \frac{2}{3}\frac{G_F^2 M_K^5}{2^8\pi^3} \beta_\ell \lambda^{1/2}(z) \times\left\{ \left|f_V\right|^2 2 \frac{\alpha^2}{16\pi^2}\lambda(z)\Big(1+2\frac{r_\ell^2}{z}\Big)
+ \left|f_S\right|^2 3\, z\beta_\ell^2 
\right\}\,,
\end{align}
where $r_{\ell,\pi} = m_{\ell,\pi}/M_K$, $\beta_\ell = \sqrt{1-4r_\ell^2/z}$, and $\lambda(z) \equiv 1 + r_\pi^4 + z^2 - 2(r_\pi^2 + z + r_\pi^2 z)$ is the K\"{a}ll\'{e}n function.
This is being measured at NA62 for both the muon and the electron mode~\cite{Boboc:2024afc}.
The presence of a scalar operators, also contributes to the forward-backward asymmetry ($A_{FB}$):
\begin{align}\label{eq:AFB}A_{\rm FB}(z) = \left. \frac{\alpha G_F^2M_K^5}{2^8\pi^4}r_\ell\,\beta_{\ell}^2(z)\lambda(z)
{\rm Re}\left(f_V^* f_S\right) \!
\middle/ \!\left(\frac{d\Gamma(z)}{dz}\right) \right.\!\!\,.
\end{align}
Due to helicity suppression, $A_{FB}$ is non-negligible only in the case of the muon mode.

The ongoing status of measurements for both these observables makes it the best prospect for the extraction of constraints on $f_S$.
Using experimental data, bounds on $f_S$ were obtained for both electron and muon channels~\cite{DAmbrosio:2024rxv}. 

In the absence of $C'_S$, these bounds can be translated into constraints on the scalar Wilson coefficient $C_S$ in Eq.~(\ref{eq:hamiltonian}) through following the relation:
\begin{align}
\vert C_S \vert = \bigg\vert \frac{2 \sqrt{2} \pi}{\lambda_t \, \alpha_{\rm em} \, f_0(q^2)} \, \frac{M_K (m_s - m_d)}{M_K^2 - M_\pi^2} \, f_S \bigg\vert\,,
\end{align}
where $f_0(q^2)$ is the scalar kaon-to-pion form factor. 
We note that the bound on the scalar form factor, $|f_s| < 7.9\times 10^{-6}$~\cite{DAmbrosio:2024rxv},  assumes it to be a constant independent of $q^2$. For consistency, we assume $f_0=1$ to be approximately constant and is justified within current experimental uncertainties.
Under these assumptions, we obtain:
\begin{equation}
|C_S| \,\lesssim\, 5.1\,,    
\end{equation}
where $|\lambda_t|$ is obtained from PDG~\cite{ParticleDataGroup:2024cfk}.
Note that this bound is mildly sensitive to the choice of $f_0$; for instance taking $f_0=1.3$, yields a stronger upper bound of $3.9$.

While the ongoing measurement at NA62 is likely to improve this, it is instructive to see the impact of other observables on $C_S$. 
This not only highlights the synergy between different measurements but also underscores the importance of observables proposed for future facilities.

\section{The near future with $K_{L,S}\to \mu^+\mu^-$}
The muonic decay mode of the two mass eigenstates of kaons are in different stages of their experimental measurements. While BR$(K_L\rightarrow \mu^+\mu^-)=(6.84\pm 0.11)\times 10^{-9}$ is measured \cite{ParticleDataGroup:2020ssz}, the existing upper bound BR$(K_S\rightarrow \mu^+\mu^-)<2.1\times 10^{-10}$ (at $~90\%$ C.L), is two orders of magnitude away from the SM prediction. The latter does not receive significant enhancements even in the presence of $(V-A)\otimes(V-A)$ type NP operators. In contrast, scalars operators can leave a significant imprint on BR$(K_S\rightarrow \mu^+\mu^-)$, thus making it a sensitive channel to probe these operators. In general, the branching fractions for  $K_{L,S}\rightarrow \mu^+\mu^-$ in the presence of (pseudo)scalar operators is given as~\cite{DAmbrosio:1986zin,Ecker:1991ru,GomezDumm:1998gw,Knecht:1999gb, Isidori:2003ts, DAmbrosio:2017klp, Mescia:2006jd,Quigg:1968zz,Martin:1970ai,Savage:1992ac,DAmbrosio:1992zqm,DAmbrosio:1996lam,Valencia:1997xe,DAmbrosio:1997eof,Cirigliano:2011ny,Colangelo:2016ruc,Dery:2021mct,Brod:2022khx,Zhao:2022pbs,Christ:2020bzb,Chobanova:2017rkj}: 

\begin{align}
 \label{eq:brKLmumuComplete}
&{\rm BR}(K^0_{L} \to \mu^+ \mu^- ) =  \tau_{L}  \frac{ f_K^2 m_K^3 \beta_{\mu,K}} { 16 \pi} \left( \frac{G_F \alpha_e}{\sqrt{2}\pi}\right)^2 
\times \Bigg\{ \beta_{\mu,K}^2\left|  \frac{ m_K}{m_s +m_d}\mbox{Im}\big(\lambda_t C_{S}\big)  \right|^2  \nonumber\\[-4pt]
&\qquad\qquad+\Bigg| \frac{\sqrt{2} \pi}{G_F \alpha_e} N_{L}^{\rm LD} - \frac{2m_\mu}{m_K} \mbox{Re}\!\left(-\lambda_c\frac{ Y_c}{s_W^2} + \lambda_t C_{10}\right) - \frac{ m_K}{m_s +m_d}\mbox{Re}\big(\lambda_t C_{P}\big)  \Bigg|^2
 \Bigg\}\,,\\[4pt]
% \end{align}
% \begin{align}
 \label{eq:brKSmumuComplete}
&{\rm BR}(K^0_{S} \to \mu^+ \mu^- ) =  \tau_{S}  \frac{ f_K^2 m_K^3 \beta_{\mu,K}} { 16 \pi} \left( \frac{G_F \alpha_e}{\sqrt{2}\pi}\right)^2 
\times \Bigg\{ \beta_{\mu,K}^2\left| \frac{\sqrt{2} \pi}{G_F \alpha_e} N_{S}^{\rm LD} - \frac{ m_K}{m_s +m_d}\mbox{Re}\big(\lambda_t C_{S}\big)  \right|^2  \nonumber\\[-4pt]
&\qquad\qquad+\Bigg| \frac{2m_\mu}{m_K} \mbox{Im}\!\left(-\lambda_c\frac{ Y_c}{s_W^2} + \lambda_t C_{10}\right) + \frac{ m_K}{m_s +m_d}\mbox{Im}\big(\lambda_t C_{P}\big)  \Bigg|^2
 \Bigg\}\,,
\end{align}
with 
\begin{align}
\beta_{\mu,K} &= \sqrt{1-4m_\mu^2/M_K^2}\,,
\end{align}
and $Y_c$ denotes the charm-loop contribution (calculated at NNLO in QCD~\cite{Gorbahn:2006bm}). The terms $N_{S,L}^{\rm LD}$ describe the long-distance contributions, taken from~\cite{Chobanova:2017rkj} and based on the analyses in~\cite{Ecker:1991ru, Isidori:2003ts, DAmbrosio:2017klp, Mescia:2006jd}.

The presence of any BSM operator  generally has a differing effect on the branching fraction of either $K_{L}$ or $K_S$ into the two-muon mode. This can be understood by the consideration of the general from of amplitude for $K^0\rightarrow \ell^+ \ell^-$ \cite{Isidori:2003ts}:
\begin{equation}
    \mathcal{M}(K^0\rightarrow \ell^+ \ell^-)=\bar u_l(iA+B\gamma_5)v_l\,.
\end{equation}
In the SM, $K_L\rightarrow \mu^+\mu^-$ will receive contribution primarily from the $B$ part of the amplitude. On the other hand, $K_S\rightarrow \mu^+\mu^-$ receives a combination of both, with the dominant coming from the $A$ part of the amplitude.
It is important to note, while the $A$ part is CP-even, the $B$ part is CP-odd.
Thus, depending on the nature of the BSM extension they can receive  contrasting enhancements over their SM predictions. When we consider a SM type short distance (SD) extension \textit{i.e.} $(V-A)\otimes(V-A)$ type \cite{DAmbrosio:2022kvb}, it contributes only to the $B$ part of the amplitude. Thus LD part of BR($K_L\rightarrow \mu^+\mu^-$) interferes constructively with the SD part at the amplitude level, resulting in a substantial enhancement over the SM expectation for a reasonable choice of the Wilson coefficient. However, there is no such interference for BR($K_S\rightarrow \mu^+\mu^-$) which results in the mode being less sensitive to the presence of SM-type NP operators. This serves as one of the strongest motivations to consider scalar operators; as seen in Eq.~(\ref{eq:brKSmumuComplete}), the scalar WC, $C_S$ interferes constructively with $N^{\rm LD}_S$. On the other hand, BR($K_L\rightarrow \mu^+\mu^-$) receives a similar enhancement only due to the interference of the corresponding long distance contribution with $C_P$.

The appearance of two free parameters, $C_S$ and $C_P$, may imply that the two decay modes are largely uncorrelated in the presence of only (pseudo)scalar operators. However, the framework can be refined 
by the consideration of the following two scenarios:
\begin{itemize}
\item A) Minimal models with the extended Higgs sector such as 2HDM, only have additional scalars in their spectrum and consequently motivate the choice: $C_P = 0$. Thus, $ C_S$ is the only free parameter.
\item B) In the scenarios where  both operators are relevant, SMEFT predicts $ C_S =- C_P$ \cite{Alonso:2014csa}.
\end{itemize}
Thus, the entire analysis can be streamlined with the consideration of a single free parameter that connects all the observables.
Within the simplified model under consideration, these scenarios represent  two independent possibilities that can illustrate the impact of scalar operators on the observables in kaon physics. In the following, we present a clinical illustration of the nature of this impact on the different observables.

However, the consideration of the two observables within the realm of Scenarios~A and~B has interesting implications. For either case, the existing measurement of BR$(K_L\rightarrow \mu^+\mu^-)$ is a driving factor in ascertaining the allowed range of the NP Wilson coefficient.
Thus, in addition to studying the impact of these operators on the individual observables, it will also illustrate the  patterns of correlations between them.

\subsection{Scenario A}
\begin{figure}[t!]
\begin{center}
\includegraphics[width=0.48\textwidth]{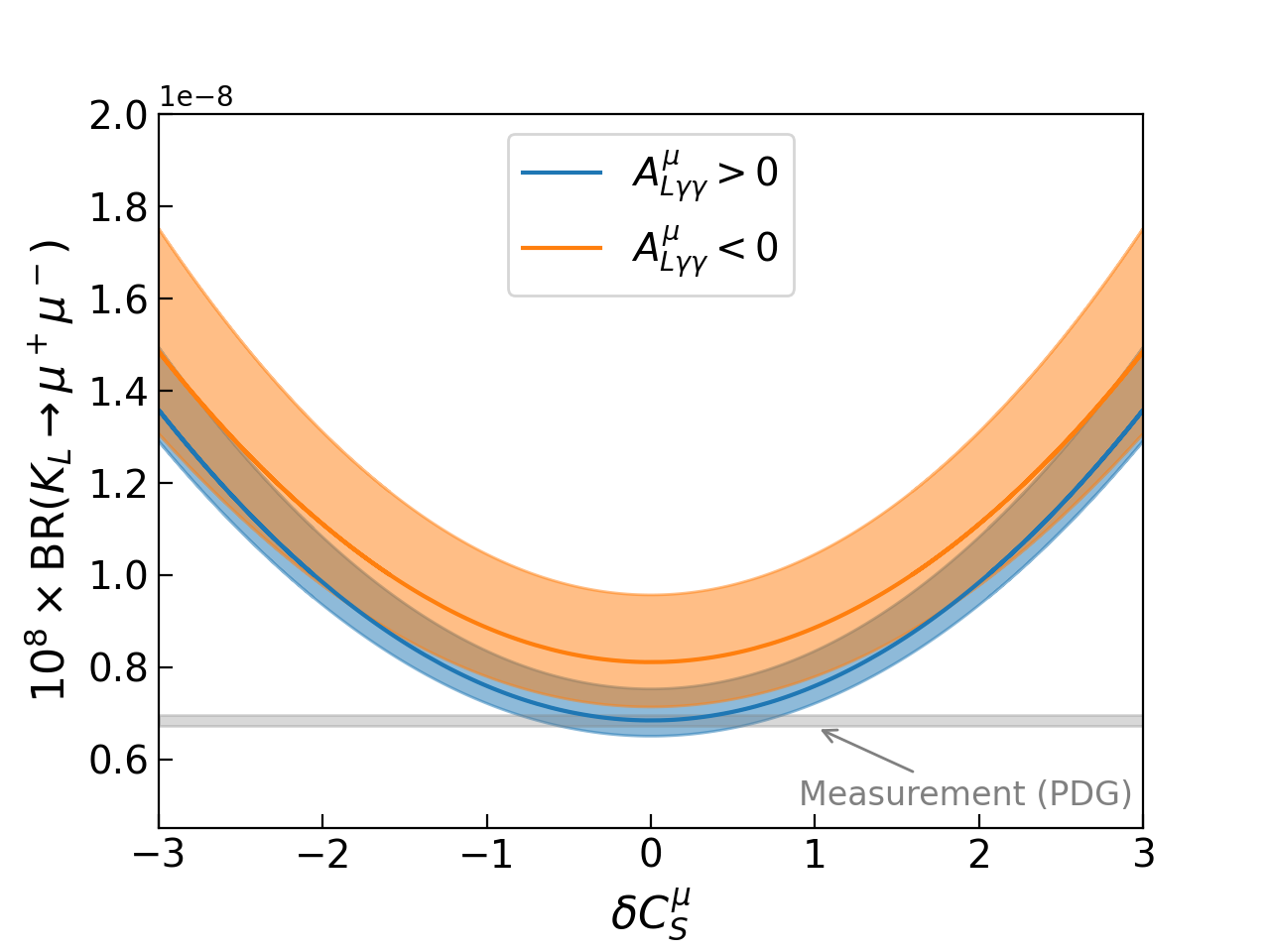}\quad  \includegraphics[width=0.49\textwidth]{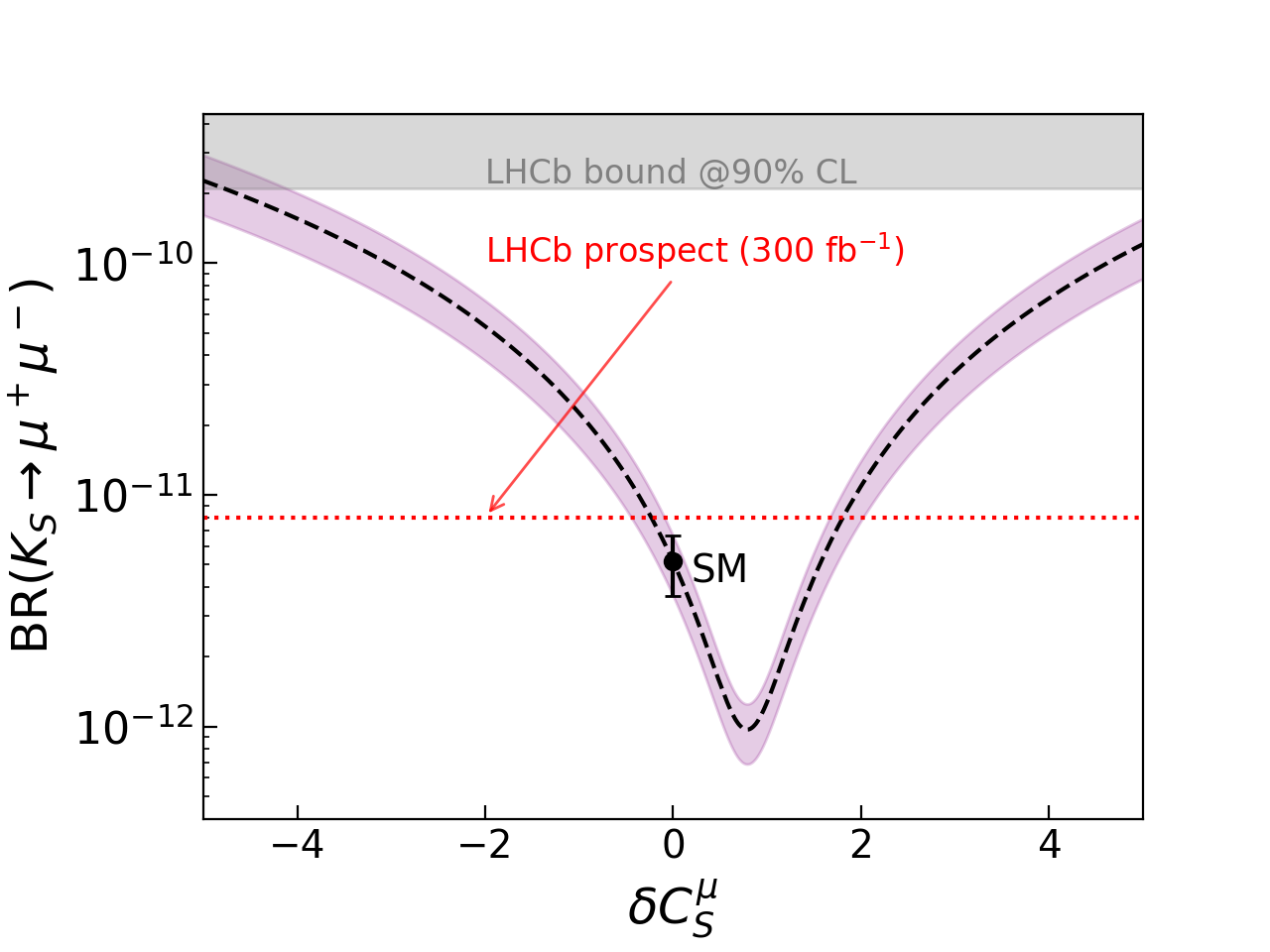}
\caption{BR($K_L\to \mu^+\mu^-$) with both sign for LD on the left and only LD:+ on the right.
\\
\label{fig:BR_KLmumu_CQ1}}
\end{center}
\end{figure}

With the assumption $\delta C_P = 0 $, we consider the effects of $\delta C_S $ on both the observables.
For the branching ratio $\mathrm{BR}(K_L \to \mu^+\mu^-)$, we consider both possible signs of the long-distance contribution (see~\cite{Hoferichter:2023wiy} for a recent evaluation of LD effects in $K_L \to \ell^+ \ell^-$).
Figure~\ref{fig:BR_KLmumu_CQ1} illustrates the effect of  a non-zero  $\delta C_S $ on BR$(K_L\rightarrow \mu^+\mu^-)$ (left) and BR$(K_S\rightarrow \mu^+\mu^-)$ (right).
The grey band on the left plot indicates the experimental measurement along with the corresponding uncertainties.
As expected, given the precise experimental determination of $\mathrm{BR}(K_L \to \mu^+\mu^-)$, this decay imposes strong constraints on the scalar coefficient $\delta C_S$, with $|\delta C_S| \lesssim 1$ at the $1\sigma$ level. By contrast, the current experimental upper bound on $\mathrm{BR}(K_S \to \mu^+\mu^-)$ is about two orders of magnitude above the SM prediction, leading to significantly weaker constraints on $\delta C_S$ from this channel. Nevertheless, it is worth noting that even at the present level, the resulting bounds are compatible with those obtained from $K^+ \to \pi^+ \mu^+ \mu^-$. Looking ahead, the projected measurement of $\mathrm{BR}(K_S \to \mu^+\mu^-)$ at the LHCb upgrade with $300~\mathrm{fb}^{-1}$ is promising, and could allow for a more stringent probe of $\delta C_S$, since sizeable deviations from the SM expectation may arise even for relatively small values of this coefficient.

%%%%%%%%%%%%%%%%%%%%%%%%%%%
In addition to looking at the individual dependency of either decay mode on $\delta C_S$, it is also instructive  to understand how they vary simultaneously with the change in $\delta C_S$.

\begin{figure}[t!]
\begin{center}
\includegraphics[width=0.49\textwidth]{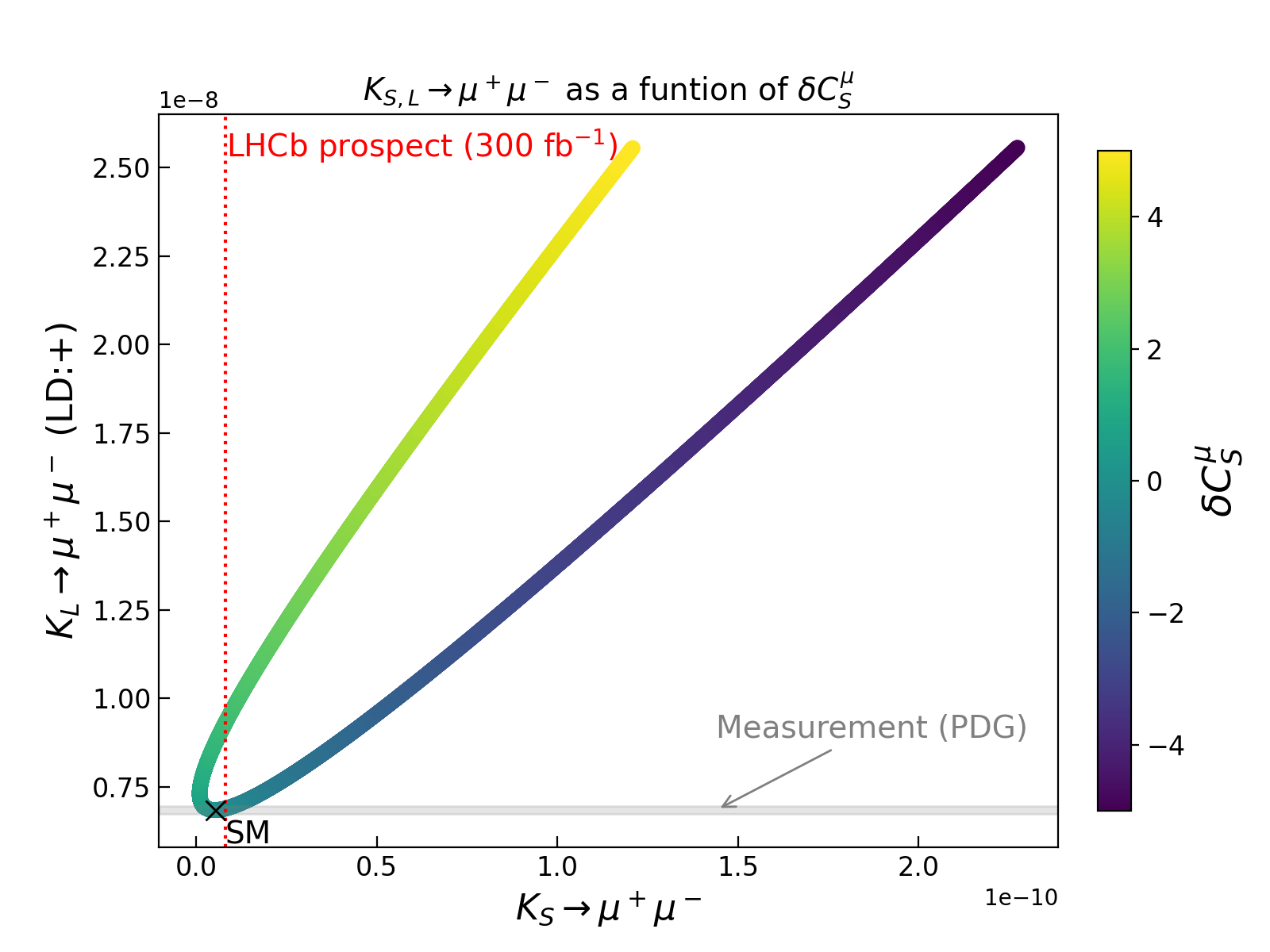}
\includegraphics[width=0.49\textwidth]{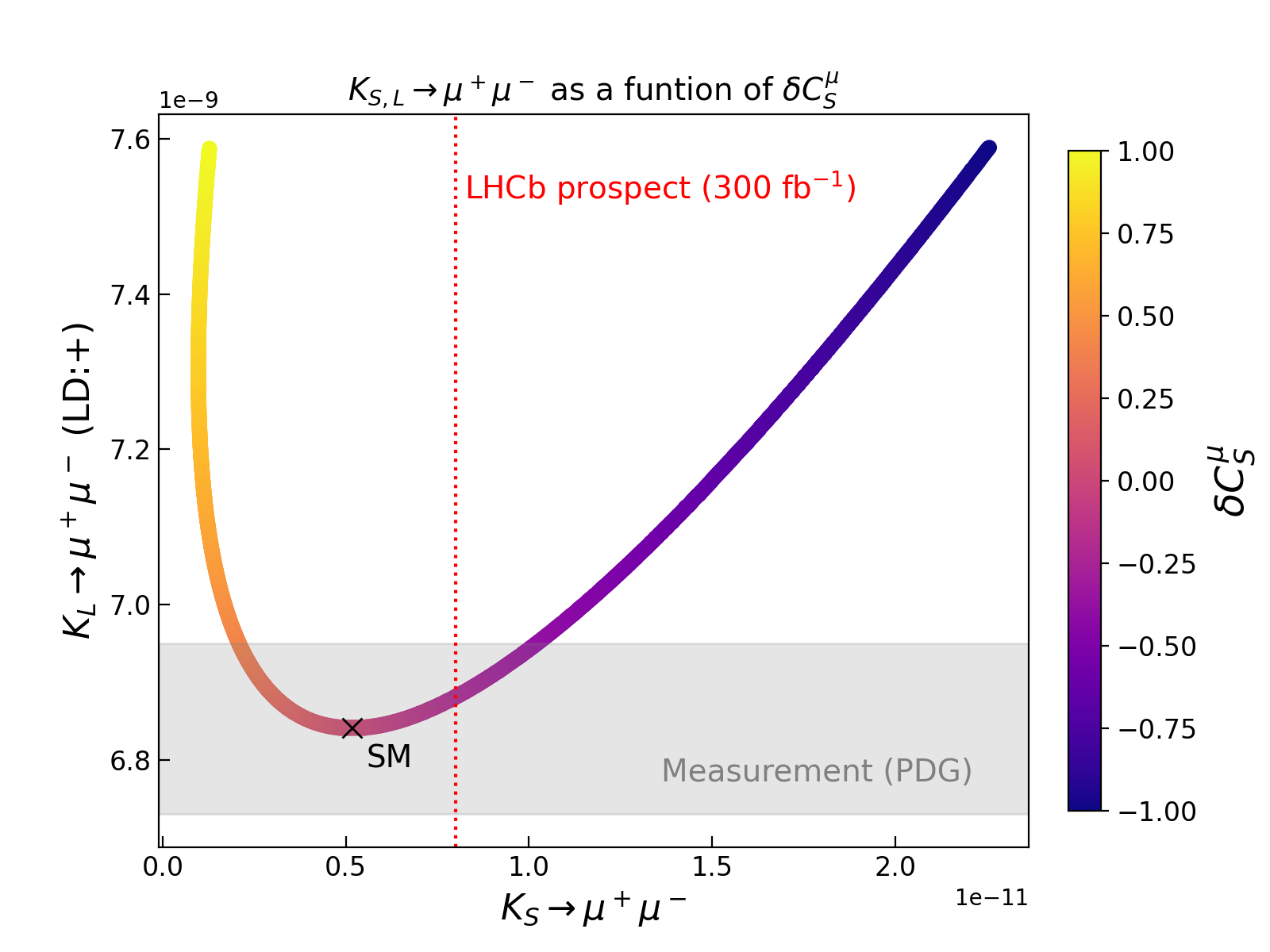}
\caption{BR($K_L\to \mu^+\mu^-$) vs 
BR($K_S\to \mu^+\mu^-$) as a function of 
$\delta C_{S}^\mu$. Zoomed version on the right.
\label{fig:BR_KLmumu_vs_KSmumu_CQ1}}
\end{center}
\end{figure}
This reveals deeper correlations between the two decay modes and is particularly useful where a measurement in one can lead to possible predictions for the other. Figure~\ref{fig:BR_KLmumu_vs_KSmumu_CQ1} underscores this correlation by means of a two dimensional plot: The left plots corresponds to the entire scanning range of $\delta C_S$, while the right plot zooms in on the parameter space of $\delta C_S$ that is permitted by the existing measurement of BR$(K_L\rightarrow \mu^+\mu^-)$. The left plot clearly reveals that even a minor deviation away from zero for $\delta C_S$, that is consistent with the experimental measurement of BR$(K_L\rightarrow \mu^+\mu^-)$, leads to a sizeable contribution to BR$(K_S\rightarrow \mu^+\mu^-)$. In particular, it can be probed with the existing projection for LHCb at $300$ fb$^{-1}$. This correlation, while intriguing, does not account for the theoretical errors for the decay modes. The existing errors, in particular for the LD computation of BR$(K_L\rightarrow \mu^+\mu^-)$, have a tendency to largely dilute this correlation. It is noteworthy that this pattern of correlation begins to re-emerge when the errors are reduced by about $30\%$. This lends a strong motivation for a better understanding of the LD contributions.

\subsection{Scenario B}
\begin{figure}[h!]
\begin{center}
\includegraphics[width=0.49\textwidth]{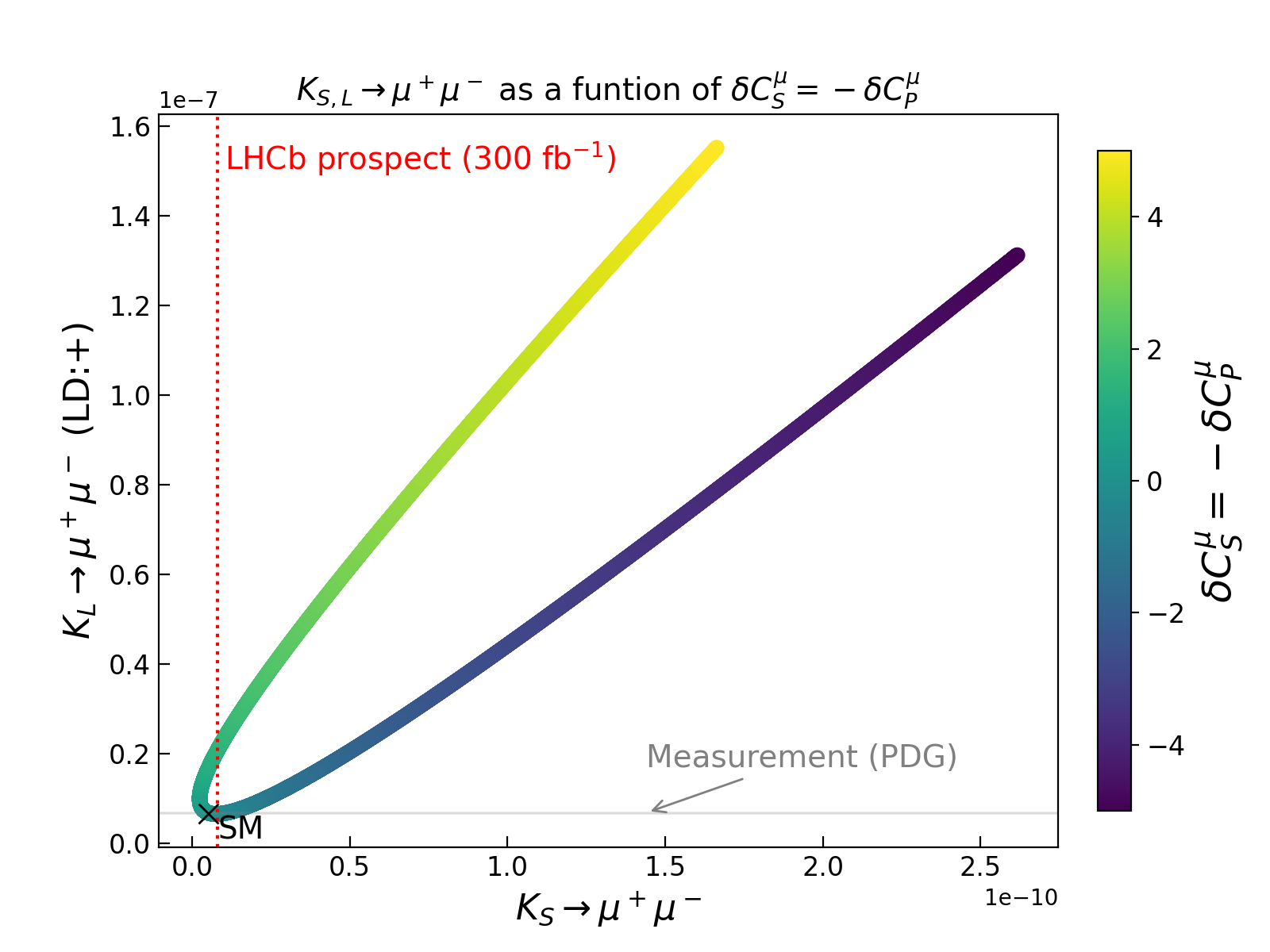}
\includegraphics[width=0.49\textwidth]{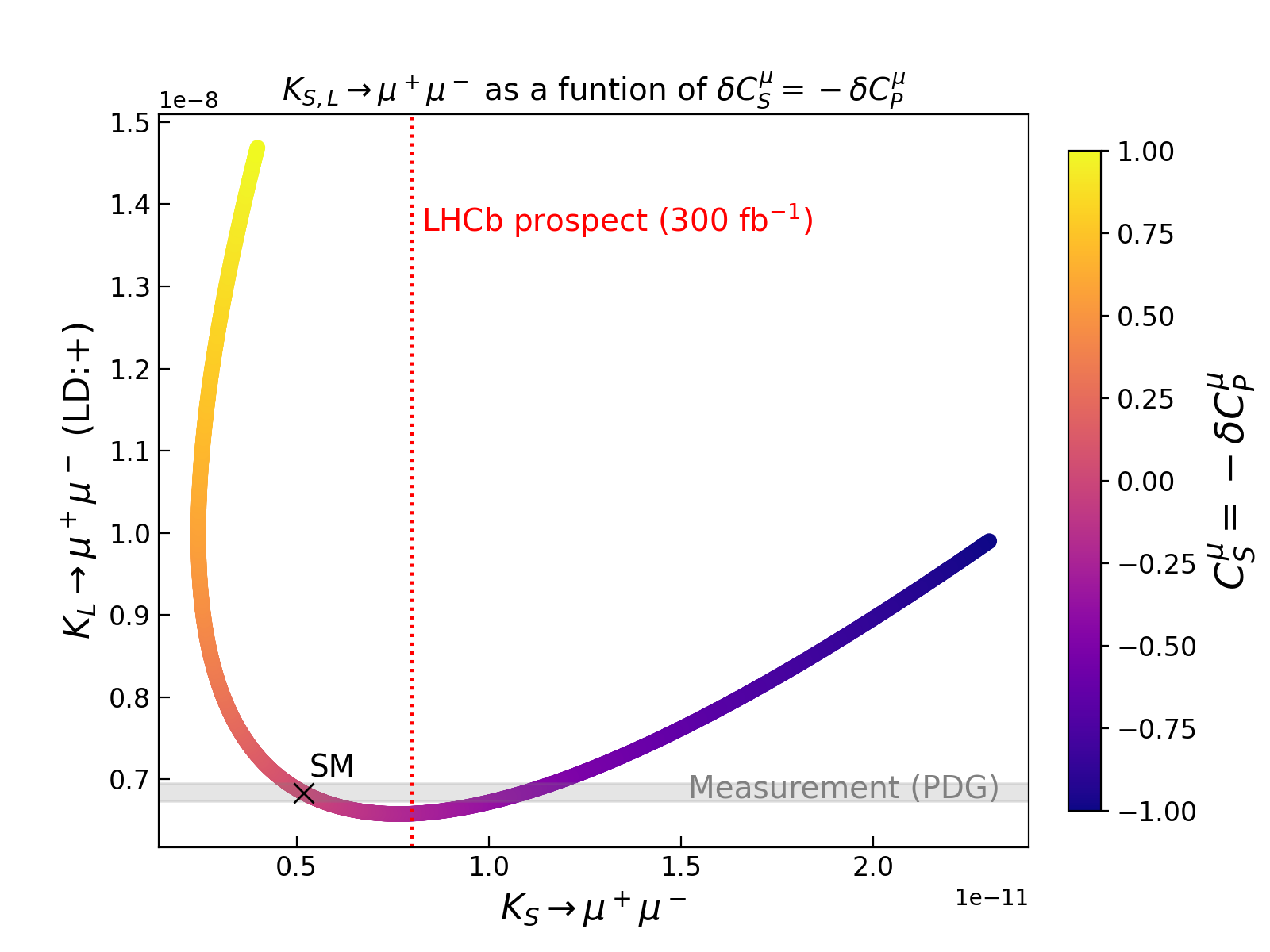}
\caption{BR($K_L\to \mu^+\mu^-$) vs 
BR($K_S\to \mu^+\mu^-$) as a function of 
$\delta C_{S}^\mu = -\delta C_{P}^\mu$. Zoomed version on the right.
\label{fig:BR_KLmumu_vs_KSmumu_CQ1_eq_mCQ2}}
\end{center}
\end{figure}

Under this approximation, both scalars and pseudoscalars contributions are at play, with $\delta C_S=-\delta C_P$. A significant deviation from their respective SM expectation is observed for both the decay modes, as shown in Fig.~\ref{fig:BR_KLmumu_vs_KSmumu_CQ1_eq_mCQ2}. In particular, for BR$(K_L\rightarrow \mu^+\mu^-)$, a sharper variation with $\delta C_S =-\delta C_P$ is seen for both signs of the LD contribution. This is is even more constraining than what was seen for this decay mode in Fig.~\ref{fig:BR_KLmumu_vs_KSmumu_CQ1}. On the other hand,  BR$(K_S\rightarrow \mu^+\mu^-)$ roughly retains the same level of sensitivity to the NP Wilson coefficient. The overall difference from Scenario A can be reconciled by considering equations \ref{eq:brKLmumuComplete} and \ref{eq:brKSmumuComplete}: the term proportional to $\delta C_P $ and  $\delta C_S$ interferes at the amplitude level with the corresponding LD part of BR($K_L\to \mu^+\mu^-$) and BR($K_S\to \mu^+\mu^-$), respectively.
Thus, Scenario B is associated with an additional  interference-level modification to the BR$(K_L\rightarrow \mu^+\mu^-)$. As a result, the values of the $\delta C_S=-\delta C_P$ that are consistent with its experimental measurement are more constrained compared to Scenario A.

We also consider the correlation between the two observables, which is shown in Fig. \ref{fig:BR_KLmumu_vs_KSmumu_CQ1_eq_mCQ2}. The left plot depicts the inter-dependency for a broader range of $\delta C_S$ and the right plot zooms in on the region that is consistent with the experimental measurement of BR($K_L\to \mu^+\mu^-$). Compared to Scenario A, the possible modification to BR($K_S\to \mu^+\mu^-$) is milder. Similar to the corresponding plots in Scenario A, the inclusion of the theoretical uncertainties would significantly diminish the effects of the correlation, thus underscoring the need to improve the theoretical computations of these decay modes.
\begin{figure}[t!]
\begin{center}
\includegraphics[width=0.48\textwidth]{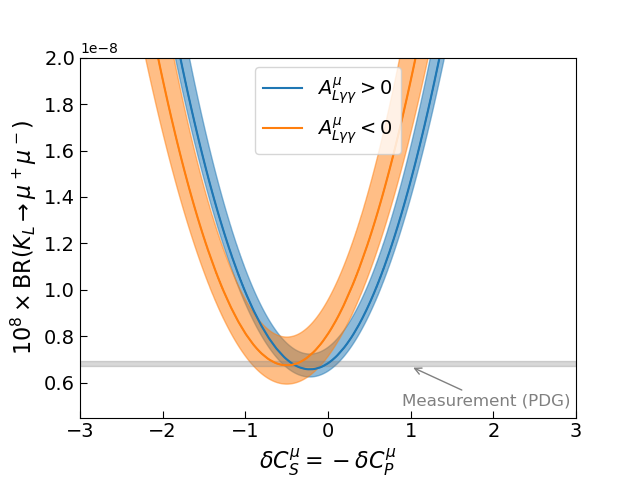}\quad \includegraphics[width=0.49\textwidth]{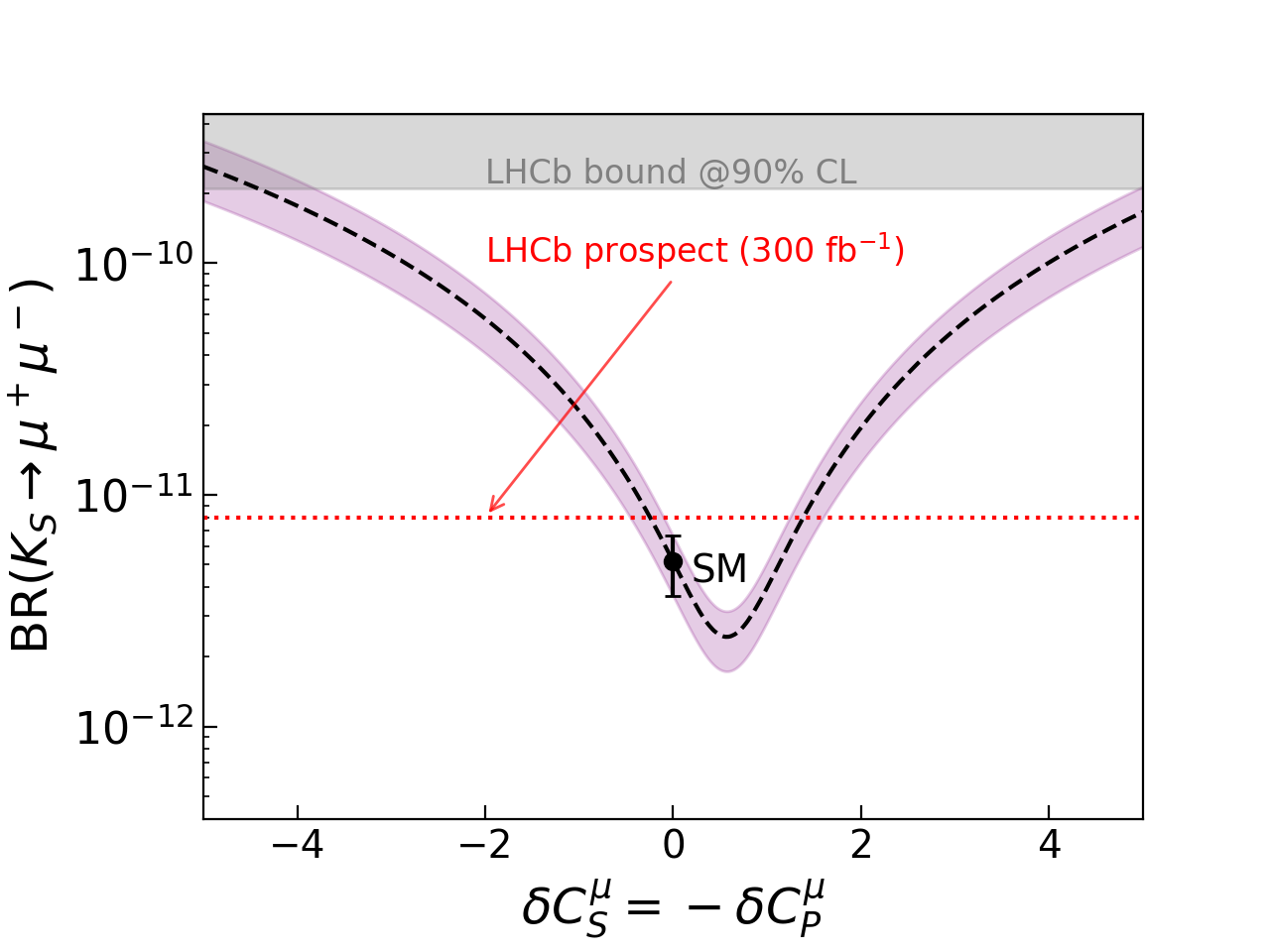}
\caption{BR($K_L\to \mu^+\mu^-$) with both sign for LD on the left and only LD:+ on the right. 
\\
\label{fig:BR_KLmumu_CQ1_eq_mCQ2}}
\end{center}
\end{figure}
%%%%%%%%%%%%%%%%%%%%%%%%%%%
%%%%%%%%%%%%%%%%%%%%%%%%%%%

\section{Prospects with $K_L \to \pi^0\ell^+\ell^-$}
As a natural extension to observables considered thus far, the semi-leptonic decay modes  $K_{L,S}\rightarrow\pi^0 \ell^+ \ell^-$, come to the forefront.
The theoretical computation of BR($K_S\rightarrow\pi^0 \ell^+ \ell^-$) is similar to that of BR($K^+\rightarrow\pi^+ \ell^+ \ell^-$), where the dominant contribution is due  to the single virtual photon exchange. The expression for the branching fraction is expressed in terms of phenomenological parameters ($a_S,b_S$) that are determined from the branching ratio or the differential decay spectrum \cite{NA48:2001kje,NA481:2004nbc}. 
A precise SM prediction for $\mathrm{BR}(K_S \to \pi^0 \ell^+ \ell^-)$ requires an accurate theoretical determination of the parameters $a_S$ and $b_S$, which is not yet available. Thus any extraction of potential (pseudo)scalar effects  can only be by means of  three parameter fits to the differential decay spectrum, similar to that proposed for $K^+\rightarrow\pi^+\ell^+ \ell^-$ \cite{DAmbrosio:2024rxv}.

The  situation differs in a subtle way
for BR($K_L\rightarrow\pi^0 \ell^+ \ell^-$), that is proposed to be measured at KOTO-II \cite{KOTO:2025gvq}. Considering only SM-type operators, this decay receives an admixture of contributions: indirect CP-violating (CPV) due to $K^0-\bar K^0$ oscillation, direct CPV due to short distance physics and a CP-conserving piece due to the two photon exchange:
branching fraction for $K_L \to \pi^0 \ell^+ \ell^-$ can be written as~\cite{Buchalla:2003sj,Isidori:2004rb,Mescia:2006jd} (see also Refs.~\cite{Martin:1970ai,Ecker:1987qi,Donoghue:1987awa,Ecker:1987hd,Ecker:1987fm,Sehgal:1988ej,Flynn:1988gy,Cappiello:1988yg,Morozumi:1988vy,Ecker:1990in,Savage:1992ac,Cappiello:1992kk,Heiliger:1992uh,Cohen:1993ta,Buras:1994qa,DAmbrosio:1996kjn,Donoghue:1997rr,KTeV:1999gik,Murakami:1999wi,KTeV:2000amh,Diwan:2001sg,Gabbiani:2001zn,Gabbiani:2002bk,NA48:2002xke})
\begin{align}
{\rm BR}(K_L \to \pi^0 \ell^+ \ell^-) = \left( C_{\rm dir}^\ell \pm C_{\rm int}^\ell |a_S| + C_{\rm mix}^\ell |a_S|^2 + C_{\gamma\gamma}^\ell \right)\times 10^{-12}\,,
\end{align}
where the parameter $|a_S| = 1.20 \pm 0.20$ is extracted from experimental measurements of the branching fractions of $K_S \to \pi^0 e^+ e^-$ and $K_S \to \pi^0 \mu^+\mu^-$, where
\begin{align}
 w_{7V} = \frac{1}{2\pi}{\rm Im}\left[ \frac{\lambda_t}{1.407\times 10^{-4}} C_9  \right]\,,\quad
 w_{7A} = \frac{1}{2\pi}{\rm Im}\left[ \frac{\lambda_t}{1.407\times 10^{-4}} C_{10}  \right]\,.
\end{align}
The SM-type contributions are dependent on the phenomenological parameters determined from $K_S\rightarrow\pi^0 \ell^+ \ell^-$. The scalar and pseudoscalar contributions are added as follows: 
\begin{align}\label{eq:BR_KLpi0ll_PS}
 \delta {\rm BR}(K_L \to \pi^0 \ell^+ \ell^-)^{S,P} = \left(C_{\rm CPC}^{\ell,SS}+ + C_{\rm CPV}^{\ell,PP}+C_{\rm CPV-int}^{\ell,PA}  + C_{\rm CPC-int}^{\ell,S\gamma\gamma}   \right) \cdot 10^{-12}\,,
\end{align}
where the first (second) term corresponds to a CP-conserving (CP-violating) contribution arising solely from scalar (pseudoscalar) operators. The remaining terms originate from interference effects: the third term represents a direct CP-violating contribution generated by the interference between pseudoscalar and axial operators, while the fourth term corresponds to a CP-conserving contribution from the interference between the scalar operator and the two-photon exchange amplitude.
Using ~\cite{Mescia:2006jd}, the value of each component is summarised in Table~\ref{tab:PS_contributions}.
\begin{table}[h!]
\renewcommand{\arraystretch}{1.3}
\centering
\scalebox{0.85}{
\begin{tabular}{c|c|c|c|c}
 & $C_{\rm CPC}^{\ell,SS}$ & $C_{\rm CPV}^{\ell,PP}$ & $C_{\rm CPV-int}^{\ell,PA}$  & $C_{\rm CPC-int}^{\ell,S\gamma\gamma}$  \\
\hline
$\ell = e$ 
& $3.9\times 10^{-7}\,(\operatorname{Re}[y_S])^2$ 
& $3.8\times 10^{-7}\,(\operatorname{Im}[y_P])^2$ 
& $1.9\times 10^{-5}\, w_{7A}\,\operatorname{Im}[y_P]$ 
& $(1.5\pm 0.3)\times 10^{-4}\,\operatorname{Re}[y_S]$ 
\\
$\ell = \mu$ 
& $4.1\times 10^{-3}\,(\operatorname{Re}[y_S])^2$ 
& $8.5\times 10^{-3}\,(\operatorname{Im}[y_P])^2$ 
& $2.6\times 10^{-1}\, w_{7A}\,\operatorname{Im}[y_P]$ 
& $(4.0\pm 1.0)\times 10^{-2}\,\operatorname{Re}[y_S]$ 
\\
\end{tabular}
}
\caption{Scalar and pseudoscalar contributions to ${\rm BR}(K_L \to \pi^0 \ell^+\ell^-)$ according to Ref.~\cite{Mescia:2006jd}.}
\label{tab:PS_contributions}
\end{table}

The scalar piece interferes with the CP-even two photon contribution while the pseudoscalar piece interferes with the CP-odd axial vector current. 
The coefficients $y_{S,P}$
are related to the Wilson coefficients, $\delta C_{S,P}$ as:
\begin{align}
    y_{S,P} = - \frac{M_W^2}{m_s\, m_\ell}\frac{s_W^2}{2} \lambda_t \delta C_{S,P}\,.
\label{eq:wckpill}
\end{align}

Thus, by means of Eq.~(\ref{eq:wckpill}), it makes it possible to see that interesting patterns can emerge due to the presence of the (pseudo)scalar pieces.
The interplay between Table~\ref{tab:PS_contributions} and Eq.~(\ref{eq:wckpill}) is noteworthy. Although Table~\ref{tab:PS_contributions} indicates that the (pseudo)scalar contributions to the electron channel are weaker than those for the muon, this effect is offset by the appearance of $m_l$ in the denominator. The two scenarios presented below provide a quantitative visualisation of this behaviour.

\subsection{Scenario A}
As this scenario is characterised by the absence of pseudoscalars, 
the second and third term (and hence $\delta C_P$) in Eq.~(\ref{eq:BR_KLpi0ll_PS}) do not contribute. 
\begin{figure}[htb!]
\begin{center}
\includegraphics[width=0.6\textwidth]{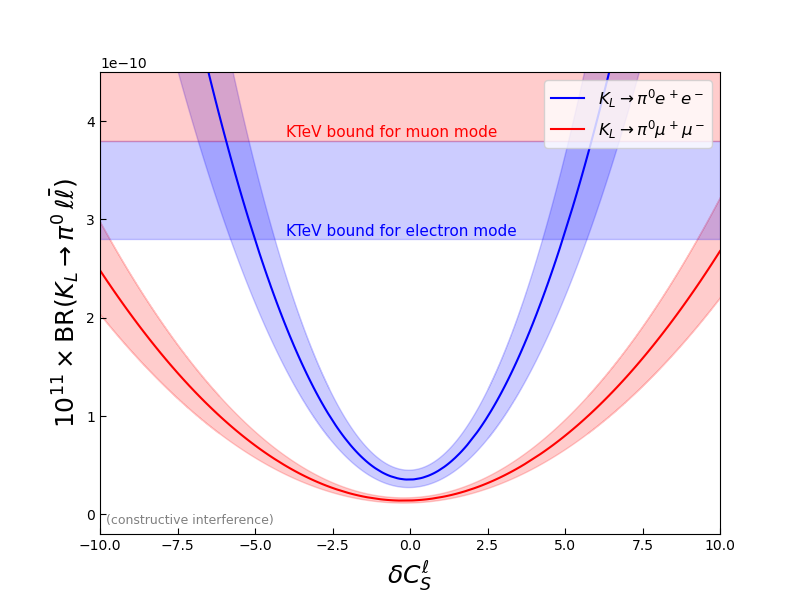}
\caption{BR($K_L\to \pi^0\ell^+\ell^-$) as a function of 
$\delta C_{S}^\mu$.
\\%\PrelErr
\label{fig:BR_KLpi0ll_CQ1}}
\end{center}
\end{figure}
The variation of the branching fraction  as a function of $\delta C_S$ is illustrated Fig. \ref{fig:BR_KLpi0ll_CQ1}. 
The blue line corresponding to the electron mode exhibits a sharper dependence on $\delta C_S$ compared to the red line of muon mode. 
The existing bound is roughly one order of magnitude larger than the SM prediction. Thus, KOTO-II, expected to reach the SM sensitivity at the $\sim 25\%$ \cite{KOTO:2025gvq, KOTO:2025uqg}, will be able to probe this operator providing complementary information to the corresponding analysis of $K_S\rightarrow \mu^+\mu^-$.

%%%%%%%%%%%%%%%%%%%%%%%%%%%

\subsection{Scenario B}
While this scenario is conceptually similar to the earlier case, the pseudoscalar pieces also contribute through the relation $\delta C_S=-\delta C_P$. As a result, for a given value of $\delta C_S$, each mode receives contributions due to the second and third piece in Eq.~(\ref{eq:BR_KLpi0ll_PS}).
\begin{figure}[htb!]
\begin{center} \includegraphics[width=0.6\textwidth]{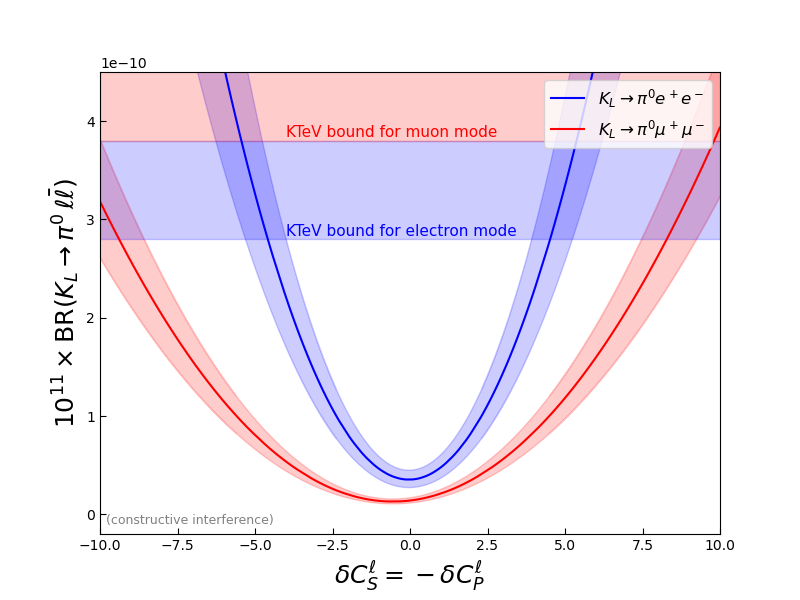}
\caption{BR($K_L\to \pi^0\ell^+\ell^-$) as a function of 
$\delta C_{S}^\mu = - \delta C_P$.
\\
\label{fig:BR_KLpi0ll_CQ1_eq_mCQ2}}
\end{center}
\end{figure}

Because of the similarity between the two scenarios, it would be very difficult to disentangle them using this particular decay mode. For either scenario, the existing upper bound on BR($K_L\to \pi^0\ell^+\ell^-$) does not have any impact on the allowed range for $\delta C_S$. However, the overall analysis with $K_L\rightarrow\pi^0\ell^+ \ell^-$ reveals two aspects: $i$) With an expected measurement at the SM level with 25$\%$ precision for either mode, it will be competitive with the future estimates on $\delta C_S$ from other modes~\cite{DAmbrosio:2024ewg}. As a consequence it complements the results from the other measurements. And $ii$) more significantly it would also open the door to study observables such as forward-backward asymmetry $A_{FB}$ at KOTO-II. 

%%%%%%%%%%%%%%%%%%%%%%%%%%%

\section{Conclusions}
The future of particle physics experiments has stimulated intense discussions in several directions, including searches for new particles and precision physics, notably in flavour physics \cite{deBlas:2025gyz}. In this context, kaon physics offers particularly strong prospects: $K_S$ decays and potentially $K_L-K_S$ interference effects can be studies at LHCb \cite{AlvesJunior:2018ldo}, while facilities at J-PARC are expected to further extend the experimental reach towards measurements of $K_L\to \pi^0 l^+l^-$~\cite{HIKE-Proposal,KOTO:2025gvq,KOTO:2025uqg,Aebischer:2025mwl}. 

In this article, we  have addressed the potential for discovering new physics in rare kaon decays their sensitivity to scalar and pseudoscalar new physics within a simplified low-energy effective framework. Restricting our attention to $(\bar{s}P_R d) (\bar{\ell} \ell)$ and $(\bar{s}P_R d) (\bar{\ell}\gamma_5 \ell)$ structures, we have analysed two representative scenarios: (A) a purely scalar case with $\delta C_P=0$, motivated by minimal extensions of the Higgs sector; and (B) a scenario with $\delta C_S=-\delta C_P$, as realised in SMEFT at dimension six.

For the dileptonic modes $K_{L,S}\to\mu^+\mu^-$, we have quantified how these operators affect both the total branching fractions and their mutual correlation. In Scenario A, $K_L\to\mu^+\mu^-$ provides the strongest constraint on $C_S$, while $K_S\to\mu^+\mu^-$ exhibits a strong dependence on the same coefficient, making it a particularly sensitive test of scalar interactions. The correlation between the two branching ratios reveals that even small deviations in $C_S$ consistent with the $K_L\to\mu^+\mu^-$ data can induce order-of-magnitude enhancements in ${\rm BR}(K_S\to\mu^+\mu^-)$. This sensitivity highlights the importance of the projected LHCb reach at ${\cal O}(300)$ fb$^{-1}$, and underlines the need for improved control over the long-distance contributions to $K_{L,S}\to\mu^+\mu^-$.

In Scenario B, where both scalar and pseudoscalar operators contribute with $C_S=-C_P$, the interference pattern in $K_L\to\mu^+\mu^-$ changes qualitatively, resulting in a stronger restriction on the allowed parameter space. The predicted correlation with $K_S\to\mu^+\mu^-$ remains (with a milder enhancement compared to scenario A), confirming that the relative phase between $C_S$ and $C_P$ plays a critical role in determining the observable effects. In both cases, theoretical progress in evaluating the long-distance amplitudes would substantially enhance the discriminating power of these observables.

The charged mode $K^+\to\pi^+\ell^+\ell^-$ provides a complementary probe. Through the differential spectrum and the forward–backward asymmetry, it constrains the same scalar coefficient $C_S$ that governs the neutral pure leptonic decays. Using existing bounds on the scalar form factor $f_S$, we have translated the experimental limits into upper bounds on $|C_S|$, currently at the level of ${\cal O}(3–5)$ units depending on the assumed hadronic input. Future improvements in the measurement of the muon channel could significantly sharpen these constraints.

Finally, we have revisited the decay $K_L\to\pi^0\ell^+\ell^-$ in the presence of (pseudo)scalar interactions. Although currently dominated by indirect and two-photon contributions (we recall that in the SM, the $\gamma\gamma$ contribution is negligible for the electron mode), this mode becomes highly sensitive to the real and imaginary parts of $C_{S,P}$ at the level targeted by KOTO-II. The complementarity of this channel with $K_{L,S}\to\mu^+\mu^-$ thus offers a coherent strategy and opportunity for probing scalar new physics in the kaon sector.

Overall, our analysis demonstrates that scalar and pseudoscalar operators can leave distinct and correlated imprints across the various rare kaon decay modes. The next generation of precision kaon experiments; NA62, the LHCb upgrade, and KOTO-II, will collectively be able to probe these effects well beyond the current limits. A sustained effort on both the experimental and theoretical fronts, in particular to reduce the uncertainties associated with long-distance contributions, will be essential to fully exploit this discovery potential.

%\clearpage
\section*{Acknowledgments}
We would like to thank Damir Becirevic for useful discussions at the Vietnam Flavour Physics Conference 2025. We also benefited from valuable discussions at the KAON 2025 workshop and with members of the NA62 Collaboration.
GD was supported in part by the INFN research initiative Exploring New Physics (ENP).
AMI acknowledges the generous support by SERB India through
project no. SRG/2022/001003.
AMI would also like to thank IP2I Lyon for the hospitality during the completion of this work. This research is funded in part by the National Research Agency (ANR) under project no. ANR-21-CE31-0002-01. 

\clearpage

\vspace{-0.5cm}
\bibliographystyle{JHEP}
\bibliography{biblio}

\end{document}